\documentstyle[12pt]{article}
\def\bq{\mbox{\boldmath$q$}}
\def\br{\mbox{\boldmath$r$}}
\def\bsig{\mbox{\boldmath$\sigma$}}

\begin{document}

\vskip 0.5 true in
\title{Pion Excess, Nuclear Correlations, and the Interpretation\\
of ($\vec p, \vec n$) Spin Transfer Experiments}
\author{Daniel S. Koltun\\
{\it Department of Physics and Astronomy}\\
{\it University of Rochester, Rochester, NY  14627-0171}}
\maketitle
\vskip 0.5 true in

\begin{abstract}
Conventional theories of nuclear interactions predict a net
increase in the distribution of virtual pions in nuclei relative to
free nucleons.  Analysis of data from several nuclear experiments
has led to claims of evidence against such a pion excess.  These
conclusions are usually based on a collective theory (RPA) of the
pions, which may be inadequate. The issue is the energy dependence of
the nuclear response, which differs for theories with strong NN correlations
from the RPA predictions. In the present paper, information about the
energy dependence is extracted from sum rules, which are calculated for
such a correlated, noncollective nuclear theory. The results lead to
much reduced sensitivity of nuclear reactions to the correlations that
are responsible for the pion excess. The primary example
is $(\vec p,\vec n)$ spin transfer, for which the expected effects are
found to be smaller than the experimental uncertainties. The analysis 
has consequences for Deep Inelastic Scattering (DIS) experiments as well.
\end{abstract}

\vfil
\eject

\section{Introduction}
This paper addresses two related questions: First, whether the analysis of data
from recent $(\vec p,\vec n)$ experiments is in disagreement with conventional 
theories of nuclear structure. Second, whether the data from these experiments
can be interpreted as evidence against the existence of pion excess in nuclei, 
as suggested by meson theories of nuclear forces. The analysis presented here
will suggest that the answer to both questions is negative. Some consequences
of this argument for the analysis of Deep Inelastic Scattering (DIS) experiments
are also explored.

The background of the questions is as follows. The idea that nuclear pions could
play a role in the EMC effect in DIS of electrons was investigated by several 
authors \cite{al,aj,af,an,am}. These pions are usually referred to as the pion
{\em excess}, excluding the pions carried by free nucleons, which are already
accounted for in the nucleon contribution to DIS. The original notion was that 
a large enhancement (of the EMC ratio) arises from collectivity of nuclei in 
`pionic' modes, as supported by RPA models of nuclear matter \cite{aj,am}. 
An alternative view \cite{af,an} is that the pion excess is 
connected to the strong two-nucleon correlations, particularly tensor and short
range correlations, in the structure of nuclear ground states. We shall take 
this second view, which associates nuclear pions with conventional nuclear
structure theory, and the meson theory of nuclear forces.

The size and origin of the pion excess could not be settled from the EMC effect
itself, which is sensitive to other physical effects as well. Previously, it 
had been proposed \cite{ar} that the collective, or 'precurser' pionic modes 
could be studied in nucleon inelastic scattering at intermediate energies. 
This is best done by spin-transfer experiments, which separate the longitudinal
and transverse response functions. A series of such experiments was 
carried out at LAMPF with polarized protons of about 500 MeV, first
for $(\vec p,\vec p\ ')$ scattering \cite{aa}, and later for $(\vec p,\vec n)$
reactions \cite{ao,ap,ab}. These last are of most direct interest for us, since
the nuclear response is purely isovector, allowing isolation of the pionic
(longitudinal) mode. 

The spin-transfer experiments are analyzed in terms of the two response 
functions, $R_L$ (spin-longitudinal) and $R_T$ (spin-transverse), which must 
be extracted from the scattering reaction data by some form of DWIA. 
(The longitudinal and transverse response functions for $(e,e')$ are different
functions from $R_L$ and $R_T$, and are obtained differently from experiment.)
The uncertainties inherent in this extraction are presumably reduced for the 
ratio $R_L/R_T$,
which is expected to be a sensitive measure of the pionic collectivity. The
predictions of RPA models of this collectivity \cite{ar,as} give enhanced values
of this ratio at moderate excitation energies of the order of the quasifree
peak. The experimental values of this ratio \cite{aa,ao,ap,ab} do not show this
collective enhancement.  This lack of enhancement 
of the $R_L/R_T$ ratio has been used as evidence for the absence of a pion 
excess. More properly, it is evidence against strong collectivity in the 
spin-longitudinal mode. However, strongly collective pionic modes have not been
found in other studies of nuclear structure, either.

Other nuclear reactions have been analyzed in terms of the possible 
contribution of nuclear pions. Dimuon production rates on nuclear targets by 
800 GeV protons \cite{au} seem to show little change (per nucleon) from $^2$H 
to Fe. This has been given as evidence against (among other things) a pion 
excess in nuclei, although the size of the expected effect is based on 
comparison with RPA predictions with sizable collectivity. The nuclear pion 
excess has also been invoked as a possible source for extra total cross section 
in $K^+$-nucleus scattering \cite{at,ag}. Here the effects are expected to be 
small, with sufficient uncertainties in the analysis to make the evidence
unclear. The connection of the nuclear pion excess to Compton scattering has 
been estimated for the deuteron and for nuclear matter with correlations 
\cite{bd}. The experimental input here is actually from the photonuclear sum 
rule, which appears to exhibit a pionic contribution.

The EMC effect, spin-transfer $(\vec p,\vec n)$ data, and the dimuon production
rates, have been taken as evidence that there is no, or at least, very little
pion excess in nuclei. Miller \cite{av}, in a comment on Alulinichev's paper
\cite{at}, argues that these experiments put a limit on the pion excess which 
is smaller than meson theory predicts \cite{af,an}. Bertsch, Frankfurt, and 
Strikman make a similar argument in an article, {\em Where are the Nuclear
Pions} \cite{ah}, concluding that meson theory fails at the relevant scales
in these experiments. Brown, Buballa, Li and Wambach accept the evidence
against conventional meson theory, but tell us in {\em Where the Nuclear Pions 
Are} \cite{ai}, how a modified theory can fit the data.

The first purpose of the present paper is to reexamine the connection between 
the nuclear response functions, $R_L$ and $R_T$, and conventional theories of 
nuclear structure, for momentum and energy transfers relevant to the recent
$(\vec p, \vec n)$ experiments. The point of view is that the ground states of 
nuclei and their excitation spectra are dominated by two-nucleon correlations 
induced by nucleon-nucleon interactions. Examples of explicit calculations
of this type by the Argonne and Urbana groups are reviewed by Benhar and 
Pandharipande \cite{ba}; we shall rely on some results of these calculations
\cite{ad} throughout this work. Consistent with this picture, we assume that 
there are no strongly collective pionic modes.

The main problem is that direct calculation of the response functions for fully
correlated nuclei is considerably more complicated than for collective modes
(in the RPA), and has not been carried out. Our method is to use information 
from sum rules for the response functions, which can be calculated (and have
been, in Ref. \cite{ad}) to construct models of the response functions 
themselves. Using these models, we find that the contributions of the 
correlations to the response functions are smaller than the experimental
uncertainties in the measured values: there is no disagreement between the
data and the models.  

Further results of our study of response functions relate to the question
of the pion excess in nuclei. For the $(\vec p, \vec n)$ data we shall see 
no disagreement with the predictions of the excess for correlated nuclei,
for reasons closely related to those discussed for the response functions.
For the EMC and dimuon production experiments, we must look at the method
by which the pion contributions are conventionally calculated, based on the
`Sullivan' process \cite{ac}. Here we shall find that kinematic constraints
reduce the contribution expected from the nuclear pions, because of the energy
structure of the response functions. It does not seem that any of these
experiments is sufficiently sensitive to the pion excess to present a 
disagreement with conventional theories of nuclear structure.

The structure of the paper is as follows. The response 
functions and the pion excess distributions are defined in Section 2. The 
energy-unweighted sum rules are discussed in Section 3, and energy-weighted 
sum rules in Section 4. The response functions are separated into quasifree
and correlation contributions; these are modeled in Section 5, using the
sum rules to constrain the functions. The results are used to interpret the 
$(\vec p,\vec n)$ data in Section 6, and the DIS structure functions in Section
7. Further discussion follows in the concluding section. 

\section{Response functions and pion excess}
It is convenient to formulate the discussion in terms of response 
functions, which are defined as follows:

\begin{equation}
R^a_\alpha (q,\omega) = \sum_f |<f| \Gamma_\alpha^a
(\mbox{\boldmath$q$})|i>|^2 \delta(\omega - E_f + E_i)
\end{equation}
where $i$ denotes the target ground state, $f$ the excited nuclear 
state, $E_i$, $E_f$ the energies of these states; \mbox{\boldmath$q$} and
$\omega$ are the momentum and energy transfers, and $\Gamma_\alpha^a$
characterizes the nuclear excitation. We assume spherical 
symmetry (unpolarized or spin-zero target) so that $R_\alpha$ depends 
only on the magnitude $q = |\mbox{\boldmath$q$}|$.

For discussion of the spin-transfer reactions we require two response 
functions:  
longitudinal $R_L^a(q,\omega)$ and transverse $R_T^a(q,\omega)$; it is these
two functions which are extracted (in principle) from the analysis of the
($\vec p, \vec n$) experiments \cite {ab}. For the range of energy and 
momentum transfers required, we assume the excitations are dominated by 
(nonrelativistic) nucleons, for which the 
$\Gamma_\alpha^a(\mbox{\boldmath$q$})$ are given by the following 
single-nucleon operators:

\begin{equation}
\Gamma_L^a(\bq) = \sum_{k=1}^A (\bsig_k \cdot \bq)\tau^a_k \quad 
{\rm exp} (i\bq \cdot \br_k),
\end{equation}
\begin{equation}
\Gamma_T^a(q) = \sqrt{1\over 2} \sum^A_{k=1} (\bsig_k \times
\bq) \tau^a_k \quad {\rm exp} (i \bq \cdot \br_k),
\end{equation}
where $\tau^a$ are the nucleon isospin (Pauli) operators, with
$a = +, -, 0$, normalized so that $\tau^2 = \sum_a \tau^{-a}\tau^a$, 
and $\mbox{\boldmath$\sigma$}$ are the Pauli spin operators, nucleons 
labeled by $k$.  For the ($\vec p,\vec n$) reaction, the charge transfer to the 
target gives $a=+$, with response functions $R^+_\alpha(q\omega)$. 

It will be easier to work in terms of the full isovector response functions,
\begin{equation}
R_\alpha(q,\omega) \equiv \sum_a R_\alpha^a(q,\omega).
\end{equation}
For the case of an isoscalar $(T=0)$ target, we may write
\begin{equation}
R_\alpha (q,\omega) = 3R^+_\alpha (q,\omega).
\end{equation}

The pion distribution in the nuclear target ($A$) is defined 
\cite{af,an} as the 
expectation of the number operator for pions of momentum \bq\  and 
charge $a$:
\begin{equation}
n_A(q) = <i|\sum_a a^+_a(\bq) a_a(\bq)|i>
\end{equation}
normalized such that the total number of pions $n_A$ is
\begin{equation}
n_A = \int {d^3q\over (2\pi)^3} n_A(q).
\end{equation}
The {\em excess} distribution is defined by the difference
\begin{equation}
\delta n_A(q) = n_A(q) - A n_N(q)
\end{equation}
where $n_N(q)$ is the pion distribution for a single nucleon (isoscalar 
average: $(n_p + n_n)/2$).

These quantities are theory dependent, and are therefore not directly 
measureable, since the virtual pion field and its
interactions are not uniquely defined. However, it makes sense to follow 
the standard pion-theoretic definitions, for comparisons with the literature,
since it is in these terms that the discussion of the nuclear pion distribution
have been carried out.
Assuming the standard form of pseudovector coupling of pions to nucleons, 
the pion distribution (6) 
may be related to the longitudinal response function $R_L$
\cite {af,ae} by
\begin{equation}
n_A(q) = {f^2F^2(q)\over 2 \varepsilon_qm^2} \int^\infty_0 d\omega 
{R_L(q,\omega)\over [\varepsilon_q + \omega]^2}.
\end{equation}
with pseudovector coupling constant $f$, form factor $F(q)$, pion 
energy $\varepsilon_q = \sqrt{q^2 +m^2}$, with $m$ the pion mass.
The excess distribution is similarly obtained
\begin{equation}
\delta n_A(q) = {f^2F^2(q)\over 2 \varepsilon_qm^2} \int^\infty_0 
d\omega{\delta R_L(q,\omega)\over [\varepsilon_q + \omega]^2}.
\end{equation}
with $\delta R_L = R_L^{(A)} - A R _L^{(N)}$, where $R_L^{(N)}$ is 
the response function for the {\em nucleon} (isoscalar average).
One often evaluates the excess distribution in the static limit, 
ignoring the nuclear excitation $\omega$ compared to $\varepsilon_q$,
in the denominator of the integral of (10). Taking $[\varepsilon_q + \omega] 
\simeq \varepsilon_q$, this becomes:
\begin{equation}
\delta n_A(q) = {f^2F^2(q)\over 2 \varepsilon^3_q m^2} 
\int^\infty_0 d\omega \delta R_L(q,\omega) .
\end{equation}
This can be evaluated in terms of a ground-state expectation value of
a two-body operator, by use of a sum rule, as discussed in the following
section.

In section 7 we shall discuss another, related function (see (46)) which has 
also been called the pion distribution function, in the context of DIS.

\section{Sum rules and correlations}
The strategy of this paper is to extract information about the response
functions from sum rules for these quantities. This will allow us to deal 
with fully interacting theories of the nuclear targets, as given by the
Argonne and and Urbana groups \cite{ba}, for which the full response
functions have not been calculated.

The first sum rule relates the integrals of the response functions over 
$\omega$ to spin-dependent correlations in the nuclear ground state. Define
the sum functions:
\begin{equation}
S_\alpha(q) \equiv {1\over 3Aq^2}\int^\infty_0 d\omega R_\alpha(q,\omega) = 
S_\alpha^{(1)}(q) + S_\alpha^{(2)}(q).
\end{equation}
The notation indicates that the sum separates into one-body and two-body parts,
which may be expressed as expectation values of one- and two-nucleon operators;
the forms follow directly from (1) - (3): 
\begin{equation}
S_L^{(1)}(q) = {1\over 3Aq^2}\sum^A_{k=1} <i|(\mbox{\boldmath$\sigma$}_k\cdot
 \bq)^2\tau_k^2|i> = 1,
\end{equation}
\begin{equation}
S_L^{(2)}(q) = {1\over 3Aq^2} \sum^A_{k\neq j} <i|(\bsig_k\cdot 
\bq)(\bsig_j\cdot \bq) \tau_k\cdot \tau_j 
{\rm exp}[i\bq\cdot(\br_k-\br_j)]|i>.
\end{equation}
\begin{equation}
S_T^{(1)}(q) = {1\over 6Aq^2}\sum^A_{k=1} <i|(\mbox{\boldmath$\sigma$}_k\times
 \bq)^2\tau_k^2|i> = 1,
\end{equation}
\begin{equation}
S_T^{(2)}(q) = {1\over 6Aq^2} \sum^A_{k\neq j} <i|(\bsig_k\times 
\bq)(\bsig_j\times \bq) \tau_k\cdot \tau_j 
{\rm exp}[i\bq\cdot(\br_k-\br_j)]|i>.
\end{equation}
Integration over the nucleon positions in (14) and (16) is implied.
The normalizations are chosen for convenience in comparison with the literature.

The sum functions $S_L(q),S_T(q)$ have been calculated for some nuclear ground 
states
\cite{ad} based on realistic nucleon-nucleon interactions and variationally 
determined correlations. Table I contains values of these functions for
$^{16}$O, for a range of $q <$ 800 MeV/$c$ \cite[Fig. 2]{ad}. 
The fact that $S_L(q) >$ 1 for
$q >$ 300 MeV/$c$, while $S_T(q) <$ 1 for all $q$ is a direct reflection of the
strong tensor correlations of nucleon pairs in the ground state.

It is useful to compare the summed functions (12) calculated for a fully
interacting nucleus to the same functions for a nuclear ground state
in the shell model approximation, where the only correlations come from
antisymmetry. Then, the resulting functions $S^{SM}(q) <$ 1. For an L-S 
closed-shell nucleus, as $^{16}$O, $S_L^{SM}(q) = S_T^{SM}(q) =S^{SM}(q)$. 
Values of this function have been calculated \cite[Fig. 2]{ag},\cite{bb}, 
and are listed in Table I. Then, define the difference functions
\begin{equation}
\Delta S_\alpha(q) = S_\alpha(q) - S^{SM}(q),
\end{equation}
which measure the effects of ground-state correlations beyond those of the
shell model. These are also given for $^{16}$O in Table I. 

The pion excess distribution can be related to the sum rule for $S_L$, in
the static approximation.
Since the nucleon contribution is subtracted from the nuclear response 
in (10), only the two-body term (14) contributes to (11).
Then the pion excess is related to 
$S_L^{(2)}$ , the two-nucleon part of the sum rule:
\begin{equation}
\delta n_A(q) = {3Aq^2f^2F^2(q)\over 2 \varepsilon^3_q m^2} 
S_L^{(2)}(q).
\end{equation}
Values of the pion excess distribution for $^{16}$O are calculated from (18) 
and the tabulated values of $S_L(q)$, using $S_L^{(2)} = S_L - 1$, 
and are listed in the last column of Table I. The 
quantity tabulated is actually $q^2\delta n_A(q)/2\pi^2A$, whose $q$-integral 
gives the pion excess per nucleon. One can see that the distribution decreases
slowly with $q$, which makes the integrated quantity quite sensitive to the
treatment of high $q$, e.g. in the form factor. In addition, the excess
distribution is negative for low $q$ and positive for higher $q$, giving
considerable cancellation in the integrated excess.
The value for $^{16}$O is $\delta n_A/A \simeq$ 0.03. (Wiringa obtains 0.027
for a similar calculation \cite{bc}.) Note that this value
corresponds to a nuclear ground state with nucleons only, as given by the
correlation function (14). Theories which include explicit $\Delta$ 
components produce values of $\delta n_A/A \simeq$ 0.08 - 0.1 \cite{af,an,bc}.
The values of $\delta n_A(q)$ are consequently more positive than those of
Table I. We return to this point in the concluding section.

\section{Sum rules and energy dependence}
In this section we combine information that can be obtained from the sum rules 
of the previous section, along with that from energy-weighted sum rules, 
to characterize the functional dependence of the response functions.  

The starting point is the separation of the response functions into
two components:
\begin{equation}
R_\alpha(q,\omega)=R_\alpha^{SM}(q,\omega)+\Delta R_\alpha(q,\omega),
\end{equation}
where $R_\alpha^{SM}$ represents the quasi-free (uncorrelated) contribution to 
$R_\alpha$, and $\Delta R_\alpha$ the contribution of the nuclear interactions.
The functional form of $R_\alpha^{SM}$ is expected to be quasifree, with a
peak in energy at $\omega\simeq q^2/2M$; we shall not require much more than
this information. For L-S nuclei, $R^{SM}$ should be independent of $\alpha$. 
We require that $R^{SM}$ obey the sum rule
\begin{equation}
S^{SM}(q) = {1\over3Aq^2}\int^\infty_0 d\omega R^{SM} (q,\omega).
\end{equation}
Similarly, if
\begin{equation}
\Delta S_\alpha(q) = {1\over3Aq^2}\int^\infty_0 d\omega \Delta R_\alpha 
(q,\omega),
\end{equation}
then (17) remains satisfied, and can be used to constrain the size of 
$\Delta R_\alpha(q,\omega)$ at fixed $q$ if the shape of the $\omega$-
distribution is known.

Some information about the shape of the $\omega$-distributions for $R_\alpha$
comes from the energy-weighted sum rule
\begin{equation}
W_\alpha(q) = {1\over3Aq^2}\int^\infty_0 d\omega \omega R_\alpha(q,\omega).
\end{equation}
By a standard double-commutator calculation, this sum can be separated into
one- and two-body parts \cite{ad}
\begin{equation}
W_\alpha(q) = {q^2\over2M} + I_\alpha(q).
\end{equation}
Some calculated values of $I_L,I_T$ \cite[Fig. 4]{ad} for $^{16}$O are given 
in Table II.

Define the energy-weighted integrals of the separated parts of (19):
\begin{equation}
W^{SM}(q) = {1\over3Aq^2}\int^\infty_0 d\omega \omega R^{SM}(q,\omega),
\end{equation}
\begin{equation}
\Delta W_\alpha(q) = {1\over3Aq^2}\int^\infty_0 d\omega \omega \Delta 
R_\alpha(q,\omega).
\end{equation}
Clearly (25) obeys the relation
\begin{equation}
\Delta W_\alpha(q) = W_\alpha(q) - W^{SM}(q).
\end{equation} 
We want to estimate $\Delta W_\alpha(q)$, which can be used in turn to determine
some characteristics of the energy distribution of $\Delta R_\alpha(q,\omega)$.
Since $W_\alpha(q)$ can be calculated from the sum rule, we need an estimate of
$W^{SM}(q)$.
First define a mean excitation energy for a given $q$
\begin{equation}
E(q) = W^{SM}(q)/S^{SM}(q),
\end{equation}
which we write in the form
\begin{equation}
E(q) = {q^2\over2M} + U(q),
\end{equation}
where $q^2/2M$ is the centroid of the quasifree energy distribution for
a Fermi gas ( with $q > 2k_F$ ), and $U(q)$ is the shift of the centroid
from purely kinetic recoil. For a symmetric quasifree peak, $E(q)$ is the
peak energy.
Then (26) can be reexpressed, using (23),(27),(28):
\begin{equation}
\Delta W_\alpha(q) = {q^2\over2M}(1 - S^{SM}(q)) + I_\alpha(q) - U(q)S^{SM}(q).
\end{equation}

We can use the position of the empirical peak in energy transfer, in the 
($\vec p, \vec n$) response function data at fixed $q$ to estimate $E(q)$.
For 240 MeV/$c < q <$ 380 MeV/$c$, the peak is found to be $\simeq$ 20 MeV
above the quasifree energy $q^2/2M$ \cite{ab}, giving $U(q) \simeq$ 20 MeV. 
Substituting this value into (29) leads the estimated values of $\Delta W_L$ 
and $\Delta W_T$ given in Table II.

A similar estimate follows from equating $W^{SM}(q) = W^{RPA}(q)$, which
may be considered to be a definition of $W^{SM}(q)$ based on
the theorem of Thouless \cite{be} for the energy weighted sum rule. 
The calculations for $^{16}$O give $W^{RPA}(q) 
\simeq 70$ MeV for $q = 350$ MeV/$c$ \cite{bf}. This gives an estimate of $U(q) 
\simeq 12$ MeV, or $\Delta W_L(q) \simeq 50$ MeV, compared to $\Delta W_L 
\simeq$ 44 MeV in Table II.

Both estimates given have the feature that $\Delta W_\alpha(q)$ is a major
fraction of $I_\alpha(q)$, or, equivalently, that $0 < U(q) << I_\alpha(q)$.
This seems reasonable, since $U(q)$ represents the average effect of 
interactions in an uncorrelated (SM) nucleus, while $I_\alpha(q)$ carries
the full effect of correlations, including tensor correlations. In fact,
for the limiting case of $q = 0$, $\Delta W_\alpha(0) = W_\alpha(0) =
I_\alpha(0)$, as is shown in Table II. This exact result follows from the
fact that $R^{SM}/3Aq^2 \to$ 0 for $q \to$ 0 for $^{16}$O, which is a
singlet in the Wigner supermultiplet scheme. (In this case $U(q)$ is not
defined, but does not influence $\Delta W_\alpha(0)$, since $S^{SM}$ vanishes.)

\section{Model estimates of $\Delta R_\alpha$}
In this section we adopt simple functional forms for $\Delta R_\alpha(q,
\omega)$, with parameters adjusted to reproduce the sum-rule quantities
$\Delta S_\alpha(q)$ and $\Delta W_\alpha(q)$, whose numerical values are
given in Tables I and II. 

For the longitudinal function we assume that $\Delta R_L(q,\omega) \geq 0$
for all $q,\omega$. Although response functions are positive (semi-)definite
(see (1)), this doesn't follow in general for the differences $\Delta 
R_\alpha$. 
However, since $\Delta S_L(q) \geq 0$ for all $q$, the assumption is plausible.
For this case it is useful to define a mean excitation energy for $\Delta R_L
(q,\omega)$ for a given $q$, as
\begin{equation}
\omega_L(q) = {\Delta W_L(q)\over\Delta S_L(q)}.
\end{equation}
Using the estimates of $\Delta W_L$ (Table II) and calculations of $\Delta S_L
(q)$ (Table I), we obtain values of $\omega_L(q)$ well in excess of 200 MeV.
These values show that there is substantial contribution to the longitudinal
response function for $\omega >$ 200 MeV, as has already been shown by
the Euclidean response calculations of Panharipande {\em et al.}\cite{ad}.
This high-$\omega$ tail, which extends well above the quasifree response,
$R^{SM}$, has important consequences, as shown in the following sections.
The following models are taken to explore the consequences of various forms of 
the tail.

For the first model we take $\Delta R_L$ to be constant over some range of
$\omega$ for fixed value of $q$, e.g., for 0 $< \omega < 2\omega_L(q)$:
\begin{equation}
\Delta R_L(q,\omega) = {\Delta S_L(q)\over2\omega_L(q)}
\Theta(2\omega_L(q) - \omega).
\end{equation}
This form has been chosen to satisfy the sum-rule differences (21) and (25),
with $\omega_L(q)$ given by (30); there are no further parameters. This model
has a symmetric distribution in $\omega$ about $\omega_L(q)$. 

The second model is chosen not to be symmetric, with exponential dependence
on $\omega$ for high values, and linear dependence for low values of $\omega$:
\begin{equation}
\Delta R_L(q,\omega) = {\Delta S_L(q)\over\beta^2}\omega \exp{(-\omega/\beta)}.
\end{equation}
This form satisfies (21) as it stands, and also (25) with $2\beta = \omega_L
(q)$. This form has its maximum value at $\omega = \beta = \omega_L(q)/2$:
\begin{equation}
\max \Delta R_L(q) = (2/e){\Delta S_L(q)\over \omega_L(q)}.
\end{equation}
The value will be $4/e$ times the constant value of (31), for any given $q$.

A model for $\Delta R_T$ must accomodate very small values of $\Delta S_T(q)$
of either sign (see Table I), together with positive values of $\Delta W_T(q)$
(see Table II). Now $\Delta R_T$ cannot be positive definite everywhere, which
requires more parameters than can be fit by the two sum rules alone. For a
simple model with a sign change at some $\omega_0$, we take
\begin{equation}
\Delta R_T(q,\omega) = {\Delta W_T(q)\over\omega_0^2}[ -\Theta(\omega_0 - 
\omega) + \Theta(\omega - \omega_0)\Theta(2\omega_0 - \omega)].
\end{equation}
This form has a positive energy tail, and a negative value for $\omega < \omega
_0$. It satisfies (25), but gives $\Delta S_T(q) = 0$, which is only 
approximately right; small adjustments to the two terms in brackets will correct
the value. One might expect that $\omega_0 \simeq \omega_L(q)$.

\section{Nuclear correlations and ($\vec p, \vec n$) data}
In this section we use the estimates of the previous section to establish the
sensitivity of the ($\vec p, \vec n$) data to the effects of correlations.
For the range 200 MeV/$c \leq q \leq$ 400 MeV/$c$, which includes the momentum
transfers of the experiments, most quantities needed for the estimates of
sensitivity vary little with $q$, so we adopt the following 'typical' values
from Tables I and II:
\begin{equation}
\Delta S_L = 0.15,
\end{equation}
\begin{equation}
\Delta W_L = 44 {\rm MeV},
\end{equation}
which gives
\begin{equation}
\omega_L = 293 {\rm MeV},
\end{equation}
using (30). For the transverse terms we take
\begin{equation}
\Delta S_T = 0.0,
\end{equation}
\begin{equation}
\Delta W_T = 39 {\rm MeV},
\end{equation}
which applies best for 200 MeV/$c \leq q \leq$ 300 MeV/$c$.

For the constant model (31), the value of $\Delta R_L$ is determined from
(35) and (37):
\begin{equation}
\Delta R_L = 2.6 \times 10^{-4} {\rm MeV}^{-1}.
\end{equation}
For the exponential model (32), the value of $\Delta R_L$ is bounded by (33):
\begin{equation}
\Delta R_L \leq 3.8 \times 10^{-4} {\rm MeV}^{-1}.
\end{equation}

For the model of $\Delta R_T$ in (34), assuming that $\omega_0 = \omega_L$,
the value for $\omega \leq \omega_L$ is given by
\begin{equation}
\Delta R_T = -4.5 \times 10^{-4} {\rm MeV}^{-1}.
\end{equation}

The natural scale with which to compare these estimates or bounds on the
values of $\Delta R_\alpha$ are the magnitudes of the response functions
$R_\alpha (q,\omega)$ extracted from the ($\vec p, \vec n$) experiments.
For the range 240 MeV/$c \leq q \leq$ 380 MeV/$c$, the data \cite[Fig 2]{ab}
peak somewhat above the quasifree recoil energy at values of
\begin{equation}
\max R_\alpha \simeq 1.2 - 1.7 \times 10^{-2} {\rm MeV}^{-1},
\end{equation}
dropping to values $\simeq 0.5 \times 10^{-2} {\rm MeV}^{-1}$ towards the
ends of the range of energy losses: $\omega \leq$ 50 MeV and $\omega \geq$
150 MeV.

As might be expected, these experimental magnitudes are of the same order
as the expected magnitude of the quasifree peak, which for a Fermi gas
(nuclear matter) is given by
\begin{equation}
\max R^{SM} = {3M\over 4k_Fq} \simeq 0.88 \times 10^{-2} {\rm MeV}^{-1}
\end{equation}
for Fermi momentum $\simeq$ 200 MeV/c, and $q \simeq$ 400 MeV/c.

Our estimates of the correlation contributions to the response functions
given by (40),(41), and (42) are then of the order of a few percent at the
peak values, and less than $10\%$ over the whole range of excitation energies,
$\omega \leq$ 150 MeV. These contributions are ${\it smaller}$ than the 
estimated uncertainties in the data quoted \cite{ab}, including counting
($\leq 10\%$), experimental systematic ($6\% - 8\%$), and model uncertainties
in extraction ($20\%, 10\%$).

The ratio $R_L/R_T$ presumably has smaller systematic and model uncertainties,
and has been used as a measure of the spin-dependent structure of the nuclear
response \cite[Fig. 1]{ab}. We can estimate this ratio, using (40) and
(42), to be
\begin{equation}
{R_L\over R_T} \simeq {{R^{SM} + \Delta R_L}\over{R^{SM} + \Delta R_T}} \simeq
{1 + {{\Delta R_L - \Delta R_T}\over{R^{SM}}}},
\end{equation}
which is of order $10\%$ in the quasifree peak region. This is again no larger 
than the experimental uncertainties.

The conclusion is that the size of the effects to be expected from short-range
and tensor interactions in the ($\vec p, \vec n$) response may not be be
large enough to be seen in the available experiments. This is in strong 
distinction to the large predicted effects of RPA models. The difference
may be traced to the broad distribution of the high-$\omega$ tails of the
response functions in the correlated nuclear theory. The breadth of these
tails limits the size of the contributions to the response functions at any
given value of $\omega$; the relation of size to breadth given by the
sum functions. In contrast to the few- to $10\%$ contributions to the 
response functions, integration over $\omega$ gives larger contributions
to the summed functions $S_\alpha$. For example, the ratio
$S_L/S_T \simeq 1.25$ for 300 MeV/$c < q <$ 400 MeV/$c$ (See Table I, or
\cite[Fig. 3]{ad}). 

\section{Pion contributions to nuclear structure functions}
The estimates made in Section 5 also have consequences for the analysis of
deep inelastic scattering (DIS) or related processes on nuclear targets.
The contribution of pions to DIS has been
treated by several authors in terms of the `Sullivan' process, in which 
virtual pions are emitted by nucleons in the target\cite{ac,aj,ak,am,aa,ai}.  
The quantity of interest in this theory is called the pion distribution 
function for the target, which may be written in terms of the longitudinal 
response function (5)\cite{aj,ak}: 
\begin{equation}
f(y) = {g^2 \over 16 \pi^2}y\int^\infty_{(My)^2}dq^2 \int^{\omega_m}_0
d\omega {F^2(t) R_L(q,\omega)\over {(t + m^2)^2}},
\end{equation}
where $t = q^2 - \omega^2$ and $y = (q_z - \omega)/M$, with $q_z$ the
longitudinal component of $\bq$ , $M$ is the nucleon mass, $g$ the pi-nucleon
pseudoscalar coupling constant (related to $f$ of (9) by $(f/m)=(g/2M)$), 
and $F(t)$ the form factor. 
The upper limit of the $\omega$-integral is given by the inequality
\begin{equation}
\omega_m \equiv q - yM \geq q_z - yM.
\end{equation}
[Note that Rees $et\ al.$ \cite{aa} use $q^2_\perp$ as the first integration 
variable, while Brown $et\ al.$ \cite{ai} use $t$. 
However, for the present discussion, it is convenient to consider $R_L$ at
fixed $q$ and $\omega$, as in (46).]

The function defined in (46) is is clearly related to the pion 
distribution defined in (9), although the two functions differ
in their kinematic variables ($y$ vs. $q$), 
in their energy denominators (see Jung and Miller \cite{ae}),
but even more significantly, in the energy cutoff (47), as we shall see. 
The pion distribution function (46) is then integrated over the momentum 
variable $y$, weighted by the pion structure function $F^{\pi}(x/y)$ to
obtain the pion contribution to the target structure function at $x$ (in the
convolution model). The interest is in the $excess$ pion contribution over that
for $A$ free nucleons, since the excess would be an aspect of nuclear 
structure (in analogy to the excess distribution defined in (8) or (10)). 

The main point of this section is the strong effect of the upper limit 
$\omega_m$ on the contribution of the nuclear response function to the
$\omega$-integral. The largest effect is to limit the contribution of
$\Delta R_L(q,\omega)$, which has a high-$\omega$ tail extending well above
$\omega_m$. By contrast, the quasifree response $R^{SM}$, or the similar
response in RPA, are less affected by the cutoff. To illustrate the effect,
we shall ignore the $\omega$-dependence of the $t$-dependent factors,
$F(t)$ and $(t + m^2)^2$, by making a {\it static approximation}: $t \to q^2$.
The remaining $\omega$-integral is now over $R_L$ alone. We define
\begin{equation}
J(q,y) = {1\over 3Aq^2} \int^{\omega_m}_0 d\omega R_L(q,\omega),
\end{equation}
which gives a measure of the pion contribution as a function of $\omega_m$.
Clearly,
\begin{equation}
J(q,y) \leq S_L(q) .
\end{equation}
For $\omega_m$ large enough to include all of $R^{SM}(q,\omega)$, we may
write (48) in the form
\begin{equation}
J(q,y) = S^{SM}(q) + f \Delta S_L(q),
\end{equation}
where $f$ is the fraction of $\Delta S_L$ included in the integral.
For the constant $\Delta R_L$ model (31), the fraction is given by
\begin{equation}
f_a = [\omega_m/2\omega_L(q)],
\end{equation}
while for the exponential $\Delta R_L$ model (32), the result (with $\beta
= \omega_L/2$) is
\begin{equation}
f_b = [1 - exp ({-2\omega_m \over\omega_L})({2\omega_m \over\omega_L} + 1)].
\end{equation}

For example, consider $q$ = 400 MeV/$c$, for which $S_L(q)$ = 1.11, which gives
the largest pion excess. With $\omega_L$ = 293 MeV as in (37), the fractions
$f_a$, $f_b$, and the values of $J_a$, $J_b$ are given in Table III, for several
values of $y$. One sees that the values of $J(q,y)$ are reduced below 1.0 for
$y \simeq 0.3$; this means that the effect of the {\it excess} pions has been 
eliminated from the integral by the kinematic cutoff, $\omega_m$.

The result is that the effect of excess pions which are associated with short-
range and tensor interactions may be sufficiently reduced by kinematic 
constraints to be inaccessible in DIS. As in the previous section, this
result follows from the extent of the tails of the response functions.
This feature has also been noticed by M. Ericson as an effect of adding
nuclear correlations to an RPA theory\cite{bg}. A similar point 
has been made by Garcia-Recio, Nieves, and Oset \cite{bh}, with regard to 
the estimates \cite{at,ag} of the pion-excess contribution to $K^+$-nucleus 
scattering, where again, a kinematic cutoff on $\omega$ limits the 
contribution of the excess pions (although in their paper, in RPA only). 

\section{Conclusions and discussion}
We have found that estimates based on a conventional theory of nuclear
structure for which the ground states and excitation spectra are dominated by
two-nucleon correlations, give small effects in the response functions
$R_L$ and $R_T$ (compared to the quasifree response), for $\omega \leq$
150 MeV. These effects are within the uncertainties reported for the data
from the recent $(\vec p,\vec n)$ experiments. Of course, there are 
uncertainties in our models used for estimates; however, we expect the order
of the effects to be well predicted. We conclude that there is
no clear disagreement between this data and this kind of nuclear structure 
theory.

The same nuclear theory predicts a pion excess distribution, which is given
for $^{16}$O in Table I, and an integrated excess per nucleon $\delta n_A/A
\simeq 0.03$. These values are for the static approximation (18), and correspond
to a contribution of the OPE potential to the binding of $^{16}$O of
$<v^{\pi}/A>\ = -30.7$ MeV \cite[Table I]{ba}. The theory includes only 
nucleons in the nuclear states. When $\Delta$-components are coupled explicitly
through a $\pi N \Delta$ vertex, the calculated excess per nucleon increases 
by about a factor of 3 \cite{af,an,bc} (again, in the static approximation).
However, the $\Delta$-contribution to the response functions would be expected
to appear at higher excitations, $\omega >$ 300 MeV, and would not contribute
to the $(\vec p,\vec n)$ data for $\omega \leq$ 150 MeV.

There is no conflict between the predicted pion excess distributions based 
on the conventional correlated nuclear theory, and the response function
data from $(\vec p,\vec n)$. That is because the former is an $\omega$-
integrated quantity, which includes all the contribution of $\Delta R_L$
at high $\omega$, while the latter only samples that high-$\omega$ tail
at fixed $\omega$, and in a range for which the quasifree response dominates.

By contrast, RPA theories which predict strong pionic ('precursor') 
collectivity do seem to be in conflict with the data. In this case, the
contribution to the pion excess will peak at values of $\omega$ in the
experimental (quasifree) region. The evidence is not against the pion 
excess, but against the collective theory of its source. It seems
likely that the choice of Landau parameter $g' \simeq$ 0.6-0.7 (which gives
such collectivity) is too low; an effective value of $g' \geq$ 0.9 seems
indicated. 

Finally, the high-$\omega$ tails for the conventional correlated 
nuclear theory suggests changes in the interpretation of the DIS and
dimuon production experiments, in terms of the pion contribution. The
present work shows that the kinematical cutoff in $\omega$ given by (47)
can reduce - even eliminate - the effect of excess pions from the
structure functions. Again, this is in contrast to the prediction of
the strongly collective theory. The point is that the pion excess as 
defined by the sum rule (18) requires integration of the energy transfer 
$\omega$ to values which may not be accessible in the DIS experiments. 
This feature has also been noticed by M. Ericson \cite{bg}, 
and by Garcia-Recio {\it et al.}\cite{bh} in a related context.

\vskip 0.3 true in
\noindent{\Large\bf Acknowledgements}
\vskip 0.2 true in

This research was supported in part by the U.S. Department of Energy under
Grant No. DE-FG02-88ER40425 with the University of Rochester. The author would
like to thank M.F. Jiang and R.B. Wiringa for unpublished results of 
calculations, and George Bertsch, Judah Eisenberg, Don Geesaman, and Gerry 
Miller for helpful comments on the manuscript.

\newpage
\begin{table}
\centering
\parbox{4in}{\caption{Sum functions calculated for $^{16}$O and pion excess
distribution, as functions of $q$}}
\vspace{.1in}
\begin{tabular}{ccccccc} \hline \hline
$q$(MeV/$c$) & $S_L$ & $S_T$ & $S^{SM}$ & $\Delta S_L$ & $\Delta S_T$ & 
${q^2\delta n}\over{2\pi^2A}$(fm) \\ \hline
0   &  0.12 & 0.10  &  0     &  0.12      &  0.10  & 0  \\
100 &  0.38 & 0.33  &  0.21  &  0.17      &  0.12  & -0.0087  \\ 
200 &  0.73 & 0.64  &  0.58  &  0.15      &  0.06  & -0.0210  \\
300 &  0.97 & 0.79  &  0.82  &  0.15      & -0.03  & -0.0044 \\
400 &  1.11 & 0.89  &  0.95  &  0.16      & -0.06  &  0.0223 \\
500 &  1.11 & 0.96  &  0.99  &  0.12      & -0.03  &  0.0264 \\
600 &  1.08 & 0.98  &  1.00  &  0.08      & -0.02  &  0.0210  \\
700 &  1.06 & 0.99  &  1.00  &  0.06      & -0.01  &  0.0162  \\
800 &  1.04 & 1.00  &  1.00  &  0.04      &  0.00  &  0.0110  \\ \hline
\hline
\end{tabular}
\end{table}

\begin{table}
\centering
\caption{Energy weighted sums calculated for $^{16}$O as functions
of $q$ }
\begin{tabular}{cccccc} \hline \hline
$q$(Mev/$c$) &  $q^2/2M$(MeV)  &  $I_L$(MeV)  &  $I_T$(MeV)  &  
$\Delta W_L$(MeV)  & $\Delta W_T$(MeV) \\ \hline

0    &   0   &  34  &  34  &  34  &  34 \\
100  &  5.3  &  37  &  37  &  37  &  37 \\
200  &   21  &  44  &  41  &  41  &  38 \\ 
300  &   48  &  52  &  47  &  44  &  39 \\
400  &   85  &  59  &  53  &  44  &  38 \\
500  &  133  &  65  &  58  &  47  &  40  \\
600  &  192  &  68  &  62  &  48  &  42  \\
700  &  260  &  69  &  64  &  49  &  44  \\
800  &  341  &  68  &  65  &  48  &  45 \\ \hline \hline
\end{tabular}
\end{table}

\begin{table}
\centering
\parbox{3.5in}{\caption{Values of $J(q,y)$ and fraction $f$ (see (50)) 
calculated for $q$ = 400 MeV/$c$, 
$\omega$ = 293 MeV, using two models of $\Delta R_L$}}
\begin{tabular}{cccccc} \hline \hline
$\ y\ $ & $\omega_m$(MeV) & $\ f_a\ $ & $\ J_a\ $ & $\ f_b\ $ & $\ J_b\ $ \\ 
\hline 
0.0 & 400 & 0.68 & 1.06 & 0.76 & 1.07 \\ 
0.1 & 306 & 0.52 & 1.03 & 0.62 & 1.04 \\ 
0.2 & 212 & 0.36 & 1.01 & 0.42 & 1.02 \\ 
0.3 & 118 & 0.20 & 0.98 & 0.19 & 0.98 \\ \hline \hline
\end{tabular}
\end{table}

\end{document}